\documentclass{ptapap}

\newcommand{\vsi}{v\,\sin i}
\newcommand{\Hm}{\langle B\rangle}
\newcommand{\Hz}{\langle B_z\rangle}
\newcommand{\Prot}{P_{\mathrm{rot}}}
\newcommand{\Feline}{Fe~{\sc ii}~$\lambda\,6149.2$\,\AA}

\author{Gautier Mathys}[ESO]
\affil[ESO]{European Southern Observatory, Alonso de Cordova 3107, Vitacura, Santiago, Chile}

\title{Sharp-Lined and Slowly Rotating Ap Stars}

\begin{document}

\maketitle

\begin{abstract}

One of the most intriguing properties of the Ap (and Bp) stars is their slow rotation, compared to the superficially normal main-sequence stars in the same temperature range. In recent years, it has emerged that several percent of the Ap stars have rotation periods exceeding one year, and that the longest rotation periods reach several centuries. Perhaps even more puzzling is the spread of the rotation periods of the Ap stars, as some of them are as short as 0.5 d. Thus, the rotation periods of these stars span 5 to 6 orders of magnitude, with no evidence for evolution besides conservation of the angular momentum during their main-sequence lifetime. Explaining how period differentiation over such a wide range is achieved in stars that are essentially at the same evolutionary stage represents a major challenge.

Consideration of the extremely slowly rotating Ap stars is of particular importance for the understanding of the origin and the evolution of the rotational properties of Ap stars as a class. We review recent progress in the knowledge of the periods and magnetic fields of the sharpest-lined and most slowly rotating Ap stars, and discuss the prospects and concerns for future progress in their study.

\end{abstract}

\section{Introduction}
\label{mathys:sec:intro}
It has long been known that, as a group, Ap stars rotate more slowly than superficially normal stars of similar temperatures \citep[e.g.,][and references therein]{1974ARA&A..12..257P}. The existence of Ap stars with very long periods, from more than 100\,d to several years, was also recognised long ago \citep{1970stro.coll..254P}. This realisation was made possible by the fact that Ap stars are oblique rotators. To first order, the magnetic fields of the vast majority of the Ap stars are dipole-like, with an axis inclined to the rotation axis. To a large extent, the magnetic field defines inhomogeneous distributions of brightness and of elemental abundances over the stellar surface, which also have a certain degree of symmetry about the magnetic axis. As a result, the changing aspect of the visible stellar hemisphere as the star rotates manifests itself observationally through strictly periodic photometric, spectroscopic and magnetic variations, whose single common period is the rotation period of the star. The anticorrelation between the variation period and the stellar equatorial velocity was convincingly illustrated by \citet{1971PASP...83..571P}. 

Thus, the rotation rates of the Ap stars can be determined directly from the observation of their periodic variations. This contrasts with most other stellar types, for which the knowledge of rotation is based on the consideration of the projected equatorial velocity,  $\vsi$, hence is limited by the ambiguity introduced by the inclination $i$ of the rotation axis to the line of sight, which is generally unknown. Furthermore, the lowest value of $\vsi$ that can be reliably determined with typical high-resolution spectrographs ($R\sim10^5$) is of the order of 3\,km\,s$^{-1}$. For typical Ap stars, this corresponds to rotation periods that do not exceed $\sim$50\,d. Hereafter, we shall refer to the Ap stars with rotation periods longer than 50\,d as the super-slowly rotating Ap (ssrAp) stars.

\section{Period distribution}
\label{mathys:sec:perdis}
The number of Ap stars for which the rotation period is known has increased dramatically in recent years through the exploitation of the results of various photometric surveys. This includes both ground-based surveys such as ASAS-3 \citep{2015A&A...581A.138B,2016AJ....152..104H} and SuperWASP \citep{2015AN....336..981B}, and space surveys such as STEREO \citep{2012MNRAS.420..757W}, Kepler \citep{2018A&A...619A..98H}, and TESS \citep{2019MNRAS.487.4695S}, for which data acquisition is still on-going and exploitation of the results is only starting. While the most recent systematic study of the distribution of the rotation periods of the Ap stars \citep{2017MNRAS.468.2745N} confirms that the majority of them have rotation periods comprised between 2 and 10 days, it also clearly shows an extended, nearly flat tail of super-slow rotators. However, photometric variation studies are poorly suited to studying this long-period tail, which as explained above, is also beyond the reach of spectroscopic line broadening studies.

With Doppler broadening below the spectroscopic resolution limit, the spectral lines of stars that have strong enough magnetic fields are observationally resolved into their magnetically split components. The wavelength separation is proportional to the mean magnetic field modulus $\Hm$, that is, the average over the visible stellar disk of the modulus of the magnetic vector, weighted by the local emergent line intensity. The value of studying the rotational variation of $\Hm$ to constrain the geometrical structure of the magnetic fields of Ap stars, which was already realised 50 years ago \citep{1970ApJ...160.1059P}, motivated the undertaking of systematic efforts to identify Ap stars with resolved magnetically split lines and study the variations of their mean magnetic field moduli over their rotation periods \citep[and references therein]{1997A&AS..123..353M,2017A&A...601A..14M}.

While the existence of ssrAp stars had already been recognised by \citet{1970stro.coll..254P},the number of such stars that were known remained small for a long time. Many more were identified in the past two decades, mostly among the newly found stars with magnetically resolved lines. The majority of the latter are genuine slow rotators (as opposed to faster rotating stars seen at a low inclination angle $i$). From consideration of the distribution of their periods, \citet{2017A&A...601A..14M} concluded that several percent of all Ap stars must have rotation periods longer than 1\,yr, that the periods of some of them must definitely be of the order of 300\,yr, and that there may even exist Ap stars with much longer rotation periods, perhaps $\sim$1000\,yr or more. Since the periods of the fastest rotating Ap stars are of the order of half a day \citep{2017MNRAS.468.2745N}, this implies that the rotation periods of the Ap stars span five to six orders of magnitude.

Understanding the origin of the huge range of rotation rates observed in the Ap stars represents a major theoretical challenge. The Ap stars loose at most a small amount of their angular momentum on the main sequence \citep{2006A&A...450..763K,2007AN....328..475H}. The evolutionary changes of their rotation periods during their main-sequence lifetimes do not exceed a factor of 2. Thus, the differentiation of their periods must mostly have been achieved before they arrive on the main sequence, or even before their progenitors become observable as magnetic Herbig Ae/Be stars \citep{2013MNRAS.429.1027A}. The way in which this happens remains elusive, and more observational constraints are needed to guide the theoretical developments. The differentiation of the rotation in Ap stars is potentially one of the main keys to the understanding of their origin and of the origin of their magnetic fields. 

Within this context, the study of the ssrAp stars is particularly valuable, not only to characterise better the long-period tail of the distribution of the rotation periods, but also because, as the most extreme examples of the slow rotation of the Ap stars, they have the potential to provide the most valuable insight into the mechanisms responsible for this generic property of the class. 

\section{The longest periods}
\label{mathys:sec:longest}
The first determination of the period of variation of an Ap star ($\alpha^2$~CVn\,=\,HD~112413, $\Prot=5.5$\,d) was achieved just over one century ago \citep{1913AN....195..159B}. Hardly more than 70 years have elapsed since the first detection of a magnetic field in a star of this type (78~Vir\,=\,HD~118022, \citealt{1947ApJ...105..105B}). It has been almost 50 years since \cite{1970PASP...82..878P} compiled the first list of Ap stars that may have long periods, raising the interest in studying such stars. The systematic study of the Ap stars with resolved magnetically split lines led to the conclusion that the ssrAp stars represent a more considerable fraction of the Ap population than was generally believed until then \citep{2017A&A...601A..14M}. The initiation of this project itself arose from the realisation a little more than 25 years ago of the potential interest of the many Ap stars with magnetically resolved lines that had not been identified yet, much less studied in detail \citep{1992A&A...256..169M}. These time scales are longer than the time bases over which observations have been obtained until now for most ssrAp stars, but still considerably shorter than the rotation periods of many of them. Accordingly, at present, only a limited fraction of the super-slow rotators have been observed over a full cycle (or more), making an accurate determination of their rotation periods possible. For the others, still in majority, only lower limits of the periods are currently known, which are of the same order as the time base of the available observations.

\begin{table*}[p!]
  \caption{Accurately known periods of ssrAp stars}
  \label{mathys:tab:ssrperiods}
\footnotesize
  \begin{tabular*}{\textwidth}[]{@{}@{\extracolsep{\fill}}rllrl}
%  \begin{tabular}{\textwidth}{rllrl}
\\
    \multicolumn{1}{c}{HD/HDE}&Other id.&Sp. type&$\Prot$ (d)&Reference\\[4pt]
    \hline\\[-4pt]
50169&BD~$-1$~1414&A3p SrCrEu&10600&\protect{\citet{2019A&A...624A..32M}}\\
9996&HR~465&B9p CrEuSi&7937&\protect{\citet{2014AstBu..69..315M}}\\
965&BD~$-$0~21&A8p SrEuCr&6030&\protect{\citet{2019A&A...629A..39M}}\\
166473&CoD~$-$37~12303&A5p SrEuCr&3836&\protect{\citet{temp1}}\\
94660&HR~4263&A0p EuCrSi&2800&\protect{\citet{2017A&A...601A..14M}}\\
187474&HR~7552&A0p EuCrSi&2345&\protect{\citet{1991A&AS...89..121M}}\\
59435&BD~$-$8~1937&A4p SrCrEu&1360&\protect{\citet{1999A&A...347..164W}}\\
18078&BD~$+$55~726&A0p SrCr&1358&\protect{\citet{2016A&A...586A..85M}}\\
93507&CPD~$-$67~1494&A0p SiCr&556&\protect{\citet{1997A&AS..123..353M}}\\
2453&BD~$+$31~59&A1p SrEuCr&518.2&\protect{\citet{2017PASP..129j4203P}}\\
51684&CoD~$-$40~2796&F0p SrEuCr&371&\protect{\citet{2017A&A...601A..14M}}\\
61468&CoD~$-$27~4341&A3p EuCr&322&\protect{\citet{2017A&A...601A..14M}}\\
110274&CoD~$-$58~4688&A0p EuCr&265.3&\protect{\citet{2008MNRAS.389..441F}}\\
69544&CD~$-$50~3225&B8p Si&236.5&\protect{\citet{2015A&A...581A.138B}}\\
188041&HR~7575&A6p SrCrEu&223.78&\protect{\citet{2017A&A...601A..14M}}\\
221568&BD~$+$57~2758&A1p SrCrEu&159.10&\protect{\citet{2017PASP..129j4203P}}\\
66821&CPD~$-$54~1483&A0p Si&154.9&\protect{\citet{2015A&A...581A.138B}}\\
116458&HR~5049&A0p EuCr&148.39&\protect{\citet{2017A&A...601A..14M}}\\
126515&BD~$+$1~2927&A2p CrSrEu&129.95&\protect{\citet{1997A&AS..123..353M}}\\
340577&BD~$+$25~4289&A3p SrCrEu&116.7&\protect{\citet{2016AJ....152..104H}}\\
263361&BD~$+$19~1468&B9p Si&88.9&\protect{\citet{2015A&A...581A.138B}}\\
149766&CD~$-$51~10384&B8p Si&87.3&\protect{\citet{2015A&A...581A.138B}}\\
$^a$&BD~$-$7~1884&SrCrEu&75.97&\protect{\citet{2016AJ....152..104H}}\\
158919&CPD~$-$65~3465&B9p Si&75.72&\protect{\citet{2015A&A...581A.138B}}\\
102797&CPD~$-$75~757&B9p Si&75.02&\protect{\citet{2002MNRAS.331...45K}}\\
8441&BD~$+$42~293&A2p Sr&69.51&\protect{\citet{2017PASP..129j4203P}}\\
5797&BD~$+$59~163&A0p CrEuSr&68.05&\protect{\citet{2018PASP..130d4202D}}\\
123335&HR~5292&Bp He wk SrTi&55.215&\protect{\citet{Hensberge:2007oz}}\\
103844&CPD~$-$75~767&A0p Si&52.95&\protect{\citet{2015A&A...581A.138B}}\\
200311&BD~$+$43~3786&B9p SiCrHg&52.0084&\protect{\citet{1997MNRAS.292..748W}}\\
256476&TYC~736-1842-1&A2p Si&51.23&\protect{\citet{2016AJ....152..104H}}\\
184471&BD~$+32$~3471&A9p SrCrEu&50.8&\protect{\citet{2012AN....333...41K}}\\
    87528&CPD~$-$62~1415&B8/9\,III$^b$&50.35&\protect{\citet{2015A&A...581A.138B}}\\[4pt]
    \hline\\[-4pt]
  \end{tabular*}
  \textit{Notes:\/} $^{(a)}$ TYC~5395-1139-1; $^{(b)}$ Spectral type from \citet{1975mcts.book.....H}.
  \end{table*}

  More specifically, to the best of our knowledge, accurate periods have been determined for 33 ssrAp stars. These stars are listed in Table~\ref{mathys:tab:ssrperiods}, in order of decreasing period length. The columns give, in order, the HD or HDE number of the star (when available), another identification, the spectral type \citep[from][unless specified otherwise]{2009A&A...498..961R}, the value of the period and the reference of the paper in which it was derived. A few long-period stars from \citet{2017MNRAS.468.2745N} were left out after critical consideration of their properties. Exclusion reasons include unreliable period value, variation period unrelated to rotation, and mistaken Ap classification.

\begin{figure}
  \resizebox{\hsize}{!}{\includegraphics{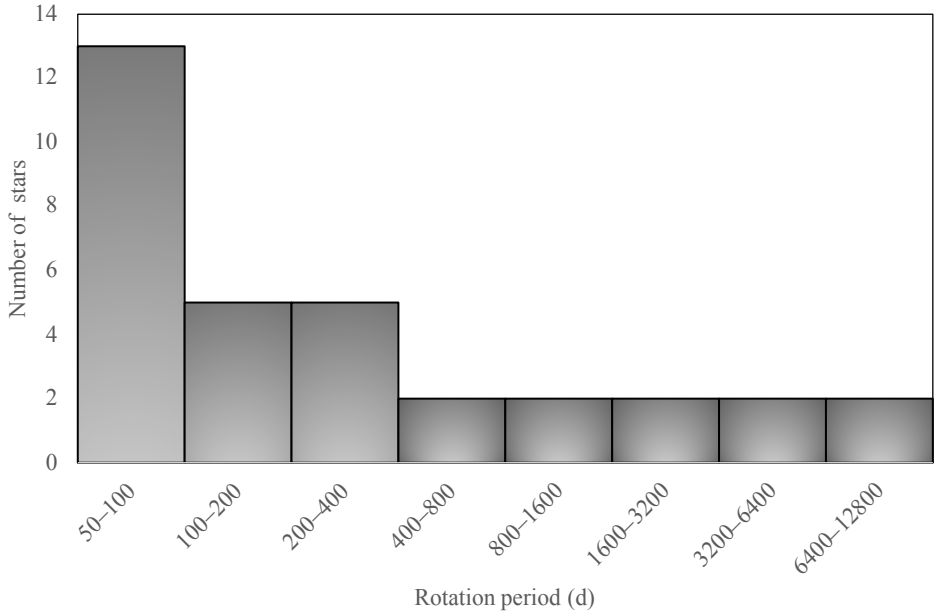}}
  \caption{Distribution of the accurately determined rotation periods of the super-slowly rotating Ap stars.}
  \label{mathys:fig:Phist}
\end{figure}

  The distribution of the rotation periods of the stars of Table~\ref{mathys:tab:ssrperiods} is shown in Fig.~\ref{mathys:fig:Phist}. At this stage, it is more representative of the current state of our knowledge than of the actual distribution of the longest rotation periods. Eight of the periods that have been accurately determined until now are longer than 1000\,d \citep{temp1}; the 29\,yr period of HD~50169 \citep{2019A&A...624A..32M} is the longest of them. This is at least ten times shorter than the period lengths  of ~300\,yr or more whose occurrence appears unescapable from statistical arguments \citep{2017A&A...601A..14M}. Among the stars for which only a lower limit of the value of the period is known until now, HD~201601 (=\,$\gamma$~Equ) has been monitored for the longest time: the available magnetic measurements span $\sim$70 years. Extrapolating from them suggests that the rotation period of this star cannot be significantly shorter than 97\,yr \citep{2016MNRAS.455.2567B}. Consideration of these numbers emphasises the incompleteness of our current knowledge of the distribution of the periods of the most slowly rotating Ap stars. Progress in this area will, by nature, be incremental, because the only way to increase the time base over which relevant observations are available is to await the passage of time. Typically, as observations covering a full rotation cycle are obtained for a growing number of stars, each increase in the value of the longest accurately determined period will be of the order of a few years. However, in the meantime, it should be possible, and valuable, to improve considerably our knowledge of the distribution of the rotation periods in the lower bins of the histogram of Fig.~\ref{mathys:fig:Phist} -- say, between 50 days and ten years.

 Of the 33 ssrAp stars whose rotation period has been accurately determined, 18 show resolved magnetically split lines. The spectral lines of another four are definitely very sharp, and do not show any hint of resolution into their magnetic components. We could not find any published measurement of the magnetic field of the remaining 11 stars, nor any high spectral resolution observation in the public observatory archives. The longest rotation period that has been derived until now for an Ap star in which magnetically resolved lines have not been observed (yet) is 236.5\,d. The 13 longest periods that have been accurately determined to this day pertain to Ap stars with resolved magnetically split lines. There are different factors that may account for this situation, either individually or in combination with each other. 

The vast majority of the known rotation periods of Ap stars have been determined through the analysis of their photometric variations. As already stated, this technique is ill-suited for long periods. The latter are best derived from the consideration of the variations of the mean magnetic field modulus, or of the mean longitudinal magnetic field ($\Hz$, the line-intensity weighted average over the visible stellar disk of the component of the magnetic vector along the line of sight), or of both. Indeed the ratio of the amplitude of variation of these field moments to the uncertainties affecting their measurements is in general much higher than the ratio of the photometric variation amplitudes to the photometric measurement errors. Furthermore the magnetic measurements are seldom affected by long-term drifts or other lacks of reproducibility that tend to occur frequently in photometric observations. This is especially true for $\Hm$ determinations, which are obtained from straightforward measurements of relative wavelength shifts of line components in standard high-resolution spectra. As $\Hz$ is diagnosed from the circular polarisation of spectral lines, the derived values may occasionally be affected by systematic instrumental effects, especially when combining observations obtained with different instruments. But for slowly rotating stars, these systematic effects tend to be limited and, often, considerably smaller than the variation amplitudes.

Naturally, the stronger the magnetic field, the easier its detection, and the higher the relative precision with which it can be measured. Therefore, using magnetic variation curves to constrain the rotation periods of the most strongly magnetic Ap stars is particularly convenient. Conversely, the scientific interest of Ap stars with spectral lines resolved into their magnetically split components and the systematic efforts that were made to identify and study such stars (see Sect.~\ref{mathys:sec:perdis}) led to the discovery of a large number of super-slow rotators, whose sometimes extremely long periods rather exceeded expectations. Undoubtedly, the combination of these two factors implies that the distribution of the magnetic fields in the ssrAp stars that are currently known is significantly biased towards the stronger ones. However, this does not rule out the possible existence of a difference between strongly and weakly magnetic Ap stars with respect to the rate of occurrence of super-slow rotation. In the course of our systematic search for Ap stars with resolved magnetically split lines, we identified a considerable number of stars whose spectral line profiles hardly differ, if at all, from the instrumental profile of the spectrograph used for their observation. In other words, the spectra of those stars do not show any significant Doppler or magnetic line broadening. If, as appears to be the case for Ap stars with magnetically resolved lines, the inclination angles of the rotation axes of these stars with respect to the line of sight are random, then the majority of them must be super-slow rotators. These stars may potentially represent a significant fraction of the group of ssrAp stars, which has been mostly overlooked until now. The distribution of their rotation periods may, or not, be similar to that of their more strongly magnetic counterparts. In particular, the question is open whether some of the weakly magnetic stars may have periods reaching several hundred years. This question is particularly relevant in relation with the apparent exclusion of very strong magnetic fields in Ap stars whose rotation periods exceed $\sim$150\,d, for which very strong statistical evidence was presented by \citet{1997A&AS..123..353M} and by \citet{2017A&A...601A..14M}.

More generally, any correlation that may exist between the rotation period and the magnetic properties of the ssrAp stars may potentially provide essential clues for the theoretical understanding of the formation and evolution of these stars. This requires the distribution of the rotation periods to be constrained across the whole range of magnetic field strengths, including the lowest ones. Together with Swetlana Hubrig, we recently recorded circularly polarised HARPS spectra of a number of the above-mentioned Ap stars with very sharp, unresolved lines that we had identified in our systematic search for Ap stars with resolved magnetically split lines. All of these stars show very definite Stokes $V$ signatures, which fully confirms the feasibility to use the variations of their mean longitudinal magnetic fields to constrain their rotation periods. We are now preparing to undertake a project to carry out systematic, multi-epoch $\Hz$ determinations for all the known Ap stars with sharp unresolved spectral lines, with a view to constraining the distribution of their rotation periods.

\section{Intermediate periods}
\label{mathys:sec:inter}
  Recently, \citet{2019ApJ...873L...5C} reported the discovery of resolved magnetically split lines in 157 Ap (and Bp) stars from medium-resolution ($R\sim22,500$) $H$-band spectra. Only three of these stars were previously known to show magnetically resolved lines in the visible: HD~14437, HD~50169, and HD~55540. This reflects in part the difference in the resolution limits between the IR study of \citeauthor{2019ApJ...873L...5C} and the bulk of the searches that were performed in the visible. Most of the latter rest on the observation of the magnetic splitting of the \Feline\ line. The Zeeman pattern of this line is a doublet, with a Land\'e factor of 2.7. For spectra of ssrAp stars recorded at $R\sim10^5$ (a typical value), \citet{1997A&AS..123..353M} argue that the components of the line can be resolved down to a field strength $\Hm\sim1.7$\,kG. 

  For simple Zeeman patterns such as pure doublets or pure triplets, the wavelength separation of the split components is proportional to the Land\'e factor of the transition and to the square of its wavelength. But for a constant multiplier, the proportionality factor is the mean magnetic field modulus of the star. For a given magnetic field strength, the wavelength dependence of the Zeeman effect represents a gain of $\sim$2.7 in the $H$ band ($\lambda\sim1.6\,\mu$m) compared to the visible ($\lambda=6150$\,\AA). However, this gain is partly negated by the lower values of the Land\'e factors of the available diagnostic lines in the IR: none of them has $g>2.0$, so that the resolution loss is $\sim$0.75 compared to the best diagnostic line in the visible. Taking into account that the resolving power of the spectra of \citet{2019ApJ...873L...5C}, $R\sim22,500$, is only $\sim$0.225 the resolving power of most of the spectra analysed by \citet{1997A&AS..123..353M} and by \citet{2017A&A...601A..14M}, $R\sim10^5$, the lower limit of magnetic line resolution in the considered $H$-band spectra must be $\sim$2.2 [$=1/(2.67\times0.75\times0.225)$] times higher than in the visible, that is, of the order of 3.8\,kG. However, this resolution is achieved for projected equatorial velocities as high as $\vsi\sim13.5$\,km\,s$^{-1}$. Thus, the type of IR observations carried out by \citet{2019ApJ...873L...5C} opens the possibility to detect resolved magnetically split lines and to determine the mean magnetic field modulus in a large number of Ap stars with rotation periods down to $\sim$12\,d.

  Thus, the Ap stars that show resolved magnetically split lines in the $H$-band study of \citet{2019ApJ...873L...5C}, on the one hand, and the Ap stars in which the \Feline\ line is resolved into its magnetic components, on the other hand, constitute two groups of stars that, although overlapping, are quite distinct from each other. To avoid ambiguity, we strongly recommend that the generic denomination ``Ap stars with resolved magnetically split lines'', without further specification, be reserved exclusively to refer to the members of the latter group, consistently with the practice that has become prevalent in the past couple of decades. The group of stars that is identified in this manner is homogeneous in the sense that the \Feline\ line is visible in the spectra of all its members, which span a temperature range that very closely coincides with that of the ``classical'' Ap stars, and that the resolution of the splitting of this Fe line is the unique and unambiguous criterion of identification of the members of the group. By contrast, different lines that are present in different stars are resolved in the $H$-band spectra analysed by \citet{2019ApJ...873L...5C}. When referring to these stars, as well as to any other star in which the reported presence of magnetically resolved lines is not based on the splitting of the \Feline\ line, one should always explicitly state the resolution criterion that was applied.

  The distinction between the two above-mentioned groups of Ap stars with resolved magnetically split lines is illustrated by the following experiment. In collaboration with Drew Chojnowski and Swetlana Hubrig, we undertook a follow-up project to observe with UVES at the Unit Telescope 2 of the ESO VLT a sample of southern stars from the list of \citet{2019ApJ...873L...5C}. Out of the 25 such stars for which a UVES spectrum had been obtained at the time of this conference, only four showed definite resolved magnetic splitting of the \Feline\ line. In the rest of the sample, the line splitting is smeared out by the Doppler broadening of the components: the projected equatorial velocity $\vsi$ of these stars is too high, so that the magnetic splitting of \Feline\ cannot be resolved. The corresponding rotation periods have not been determined and they cannot be estimated from consideration of the line widths on an individual basis, since the inclinations of the rotation axes to the line of sight are unknown. However, on a statistical basis, most of them must be comprised between $\sim$12\,d and $\sim$50\,d, as per the arguments developed above. This conclusion is reasonably consistent with the rate of occurrence of Ap stars with $12\,\mathrm{d}\le\Prot\le50$\,d in the list compiled by \citet{2017MNRAS.468.2745N}: these stars represent $\sim$70\%\ of the total number of Ap stars with accurately determined rotation periods longer than 12\,d. 

Accordingly, consideration of the group of stars identified by \citet{2019ApJ...873L...5C} is complementary to the study of the ``classical'' Ap stars with resolved magnetically split lines \citep{2017A&A...601A..14M} and opens the possibility to characterise the Ap stars with ``intermediate'' rotation periods, both in terms of the distribution of their periods and of the possible connections between their rotation and their magnetic properties. This intermediate period range is a part of the parameter space that has received little attention until now and that deserves to be further explored. 

\section{Rate of occurrence of super-slow rotation}
\label{mathys:sec:rate}
Our current knowledge of the overall rate of occurrence of the ssrAp stars is limited by the biases affecting the existing studies of the longest period stars, which were discussed in Sect.~\ref{mathys:sec:longest}. This significantly hampers theoretical developments, as a complete and accurate knowledge of the distribution of the rotation periods represents an essential constraint for the models. The TESS Mission presents an opportunity to overcome those biases, to a large extent. Indeed, an exhaustive list of Ap stars was proposed for observation by TESS during the nominal mission \citep{2019MNRAS.487.3523C}. Ultimately, not all were observed, but the selection was based on the overall priorities of the mission, not on the properties of the Ap targets: it is unbiased with respect to these properties. Thus, any inference about the longest-period Ap stars that is derived from the TESS observations is representative of their actual rate of occurrence. In particular, all stars are dealt with in the same way, regardless of the strength of their magnetic field. This ensures that weakly magnetic stars are duly included in the statistics.

As part of a project in collaboration with Don Kurtz and Daniel Holdsworth, we have started to identify the Ap stars that do not show evidence of photometric variations of rotational nature in the 27-d long data sets recorded in each of the TESS sectors. The vast majority of these stars very probably have rotation periods considerably longer than 27\,d. The main exceptions should be those stars whose rotation axis lies almost exactly along the line of sight. There should be very few such stars: under the assumption that the inclination angles of the rotation axes with respect to the line of sight are random, these angles would be less than $5^\circ$ for less than 1\% of the stars. We checked that the adopted strategy successfully identified almost all the known ssrAp stars present in the observed fields. The rare exceptions all seem to be cases where apparent variations with periods shorter than 27\,d that were detected by TESS can plausibly be attributed to contamination by the light of another, unresolved, neighbouring source. 

We shall record high-resolution spectra of all the new long-period candidates identified through the above-described search, to confirm that they are Ap stars with a low projected equatorial velocity. The resulting sample will presumably include a mix of stars with resolved magnetic split lines corresponding to a range of field strengths and of stars showing sharp spectral lines with little or no evidence of a magnetic effect. Further follow-up of these stars will be carried out to constrain their rotation periods. Ultimately, this study will enable us to characterise the distribution of the ssrAp stars both in terms of periods and in terms of magnetic field strengths. 

\section{Final remarks}
\label{mathys:sec:final}
How the differentiation of the rotation rates of the Ap stars over five to six orders of magnitude is achieved is one of the outstanding unsolved questions of stellar physics. Answering it is essential to understand the formation of these stars, and the origin of their magnetic fields. More generally, it is relevant for the knowledge of the physical processes that are at play in the formation and evolution of all the early-type stars. This relevance is emphasised by the recently evidenced similarity between the distributions of the rotation periods of the Ap (and Bp) stars, the magnetic early B-type stars and the magnetic O stars \citep{2018MNRAS.475.5144S}.

The knowledge of the distribution of the rotation periods of the Ap stars, and in particular, of its long-period tail, is essential as input for the theoretical developments aimed at explaining the evolution of rotation in these stars. In recent years, the exploitation of the results of several extensive ground- and space-based photometric surveys has led to a considerable increase in the number of well determined short periods, up to 2--3 weeks. The corresponding part of the distribution is, accordingly, very well defined. By contrast, for more slowly rotating stars, the current knowledge of the period distribution remains biased and incomplete.

Some of these shortcomings can now be addressed. In this presentation, we showed how the ``non-variable'' Ap stars in the TESS database can be exploited to constrain the rate of occurrence of ssrAp stars. In particular, considering them will allow us to identify, and to characterise, the weakly magnetic ssrAp stars, which have never been systematically studied until now. We also illustrated how the usage of medium-resolution IR spectroscopy can contribute to improving the characterisation of the stars in the intermediate period range, from $\sim$2 weeks to $\sim$2 months.

However, for the longest periods, there is a fundamental limitation that can only be overcome by the passing of time. Namely, no period can be accurately determined that is longer than the time base over which suitable observations of the star of interest have been obtained. Currently, the longest period that has been accurately determined is that of HD~50169, $\Prot=29$\,yr. Most likely, the star that will someday overturn HD~50169 as the longest-period Ap star with full coverage of the variation curve of one of its observables will only have a rotation period a few years longer. Such increments are small compared to the century-scale periods of the most slowly rotating stars. With respect to the latter, we are responsible to ensure that no gaps are left in the coverage of their variation curves. Decades or centuries may elapse before a critical phase that is missed now can be reobserved.

\bibliographystyle{ptapap}
\bibliography{mathys}

\end{document}